\documentclass[12pt,preprint]{aastex}

\newcommand{\teff}{\ensuremath{T_{\rm eff}}}
\newcommand{\logg}{\ensuremath{\log{g}}}

\newcommand{\rstar}{\ensuremath{R_\star}}

\newcommand{\rpl}{\ensuremath{R_{\rm P}}}

\newcommand{\kep}{{\em Kepler}}

\shortauthors{LATHAM ET AL.}
\shorttitle{COMPARISON OF KEPLER SINGLES AND MULTIPLES}
\slugcomment{Version 10, 19 March 2011}

\begin{document}

\title{A FIRST COMPARISON OF KEPLER PLANET CANDIDATES IN SINGLE AND MULTIPLE SYSTEMS}

\author{
David~W.~Latham\altaffilmark{1},
Jason~F.~Rowe\altaffilmark{2},
Samuel~N.~Quinn\altaffilmark{1},
Natalie~M.~Batalha\altaffilmark{3},
William~J.~Borucki\altaffilmark{2},
Timothy~M.~Brown\altaffilmark{4},
Stephen~T.~Bryson\altaffilmark{2},
Lars~A.~Buchhave\altaffilmark{5},
Douglas~A.~Caldwell\altaffilmark{6},
Joshua~A.~Carter\altaffilmark{1,20}
Jessie~L.~Christiansen\altaffilmark{6},
David~R.~Ciardi\altaffilmark{7},
William~D.~Cochran\altaffilmark{8},
Edward~W.~Dunham\altaffilmark{9},
Daniel~C.~Fabrycky\altaffilmark{10,20},
Eric~B.~Ford\altaffilmark{11},
Thomas~N.~Gautier,~III\altaffilmark{12},
Ronald~L.~Gilliland\altaffilmark{13}
Matthew~J.~Holman\altaffilmark{1},
Steve~B.~Howell\altaffilmark{14,2},
Khadeejah~A.~Ibrahim\altaffilmark{15},
Howard~Isaacson\altaffilmark{16},
Jon~M.~Jenkins\altaffilmark{6},
David~G.~Koch\altaffilmark{2},
Jack~J.~Lissauer\altaffilmark{2},
Geoffrey~W.~Marcy\altaffilmark{16},
Elisa~V.~Quintana\altaffilmark{6},
Darin~Ragozzine\altaffilmark{1},
Dimitar~Sasselov\altaffilmark{1},
Avi~Shporer\altaffilmark{4,17},
Jason~H.~Steffen\altaffilmark{18},
William~F.~Welsh\altaffilmark{19},
Bill~Wohler\altaffilmark{15}
}

\altaffiltext{1}{Harvard-Smithsonian Center for Astrophysics,
60 Garden Street, Cambridge, MA 02138, USA}
\altaffiltext{2}{NASA Ames Research Center, Moffett Field, CA 94035, USA}
\altaffiltext{3}{San Jose State University, San Jose, CA 95192, USA}
\altaffiltext{4}{Las Cumbres Observatory Global Telescope, Goleta, CA 93117, USA}
\altaffiltext{5}{Niels Bohr Institute, Copenhagen University, DK-2100 Copenhagen, Denmark}
\altaffiltext{6}{SETI Institute/NASA Ames Research Center, Moffett Field, CA 94035, USA}
\altaffiltext{7}{NASA Exoplanet Science Institute/Caltech, Pasadena, CA 91125, USA}
\altaffiltext{8}{University of Texas, Austin, TX 78712, USA}
\altaffiltext{9}{Lowell Observatory, Flagstaff, AZ 86001, USA}
\altaffiltext{10}{University of California, Santa Cruz, CA 95064, USA}
\altaffiltext{11}{University of Florida, Gainesville, FL 32111, USA}
\altaffiltext{12}{Jet Propulsion Laboratory/California Institute of Technology, Pasadena, CA 91109, USA}
\altaffiltext{13}{Space Telescope Science Institute, Baltimore, MD 21218, USA}
\altaffiltext{14}{National Optical Astronomy Observatory, Tucson, AZ 85719, USA}
\altaffiltext{15}{Orbital Sciences Corporation/NASA Ames Research Center, Moffett Field, CA 94035, USA}
\altaffiltext{16}{University of California, Berkeley, Berkeley, CA 94720, USA}
\altaffiltext{17}{University of California, Santa Barbara, CA 93106, USA}
\altaffiltext{18}{Fermilab Center for Particle Astrophysics, P.O. Box 500, Batavia, IL 60510, USA}
\altaffiltext{19}{San Diego State University, San Diego, CA 92182, USA}
\altaffiltext{20}{Hubble Fellow}

\begin{abstract}
In this letter we present an overview of the rich population of
systems with multiple candidate transiting planets found in the first
four months of \kep\ data.  The census of multiples includes 115
targets that show 2 candidate planets, 45 with 3, 8 with 4, and 1 each
with 5 and 6, for a total of 170 systems with 408 candidates.  When
compared to the 827 systems with only one candidate, the multiples
account for 17 percent of the total number of systems, and a
third of all the planet candidates.  We compare the characteristics of
candidates found in multiples with those found in singles.  False
positives due to eclipsing binaries are much less common for the
multiples, as expected.  Singles and multiples are both dominated by
planets smaller than Neptune; $69^{+2}_{-3}$\ percent for singles and
$86^{+2}_{-5}$ percent for multiples.  This result, that systems with
multiple transiting planets are less likely to include a transiting
giant planet, suggests that close-in giant planets tend to disrupt the
orbital inclinations of small planets in flat systems, or maybe even
to prevent the formation of such systems in the first place.
\end{abstract}



\section{INTRODUCTION}

Although it was anticipated that NASA's \kep\ mission could find
systems with more than one planet transiting the same host star
\citep{Koch:96,Holman:05}, the rich harvest of candidate multiples
that appeared already in the first four months of \kep\ data caught
all of us on the \kep\ Science Team by surprise.  The first
announcement of five multiples \citep{Steffen:10} was timed to
coincide with the initial public data release on 15 June 2010
\citep{Borucki:11a}.  Systems with multiple transiting planets are
rich with information that provides additional constraints on the
characteristics of the planets and even their host stars
\citep{Ragozzine:11}, as illustrated by two examples of multiple
planet systems exhibiting transit time variations that constrain the
masses of the planets: Kepler-9 with three transiting planets
\citep{Holman:10,Torres:11}, and Kepler-11 with six
\citep{Lissauer:11a}.  In this Letter we present an overview of the
full population of multiples that show transits in the first four
months of \kep\ data.  Note that we have not attempted to correct for
the probability that planetary orbits are properly aligned to show
transits, nor for the dependence of transit detectability on various
noise sources.  Instead we have chosen to compare singles with
multiples in ways that should minimize these biases.

\section{KEPLER OBJECTS OF INTEREST}

\kep\ targets that show features in their light curves that might be
due to transits are designated ``Kepler Objects of Interest'' (KOIs).
The KOI numbering convention is that the digits before the decimal
point specify a unique target, and the two digits after specify a
planet candidate, in the order that it was identified for that target.
There is no simple description of how KOIs were identified, because
the procedures evolved considerably as the data improved and the team
gained experience.  The general approach used for the identification
of KOIs is described by \citet{Borucki:11b}.  Here we present some
additional details, with special emphasis on the procedures that were
used to identify candidates in multiples.

Initially, KOIs were identified by visual inspection of light curves
for candidates identified by the \kep\ pipeline using the Transiting
Planet Search \citep[TPS;][]{Jenkins:10} on individual quarters of
data. The Data Validation \citep[DV;][]{Wu:10} reports from the
pipeline were then used to identify false positives involving centroid
motion during dimmings and also to identify additional candidates.
This effort resulted in nearly 1000 KOIs and somewhat less than 100
systems of multiple candidates.

The next major release of the \kep\ pipeline will stitch quarters
together, so that TPS and DV can work on light curves from multiple
quarters.  As a stopgap, a stand-alone tool for analyzing multiple
quarters was developed by Jason Rowe.  Starting with the calibrated
(raw) time series, sections were excised that showed instrumental
artifacts, such as gaps due to safe modes of the spacecraft and
subsequent thermal settling.  The light curves were next detrended
with a high-pass filter (to reduce sensitivity to instrumental
drifts and long-term stellar variability) and then were searched for
transits using a version of the Box Least Squares
\citep[BLS;][]{Kovacs:02} algorithm.  Multi-quarter data for nearly
180,000 targets were searched for transits (some targets were observed
for only one or two quarters).  Any event that was detected above a
3-$\sigma$ threshold was sent to routines that attempted to fit a
planetary-transit model, adopting the stellar parameters (\teff,
\logg, and \rstar) from the Kepler Input Catalog
\citep[KIC;][]{Brown:11}.  Plots of the successful fits, about 25,000
in all, were then inspected visually.  This effort led to about 600
additional KOIs, with nearly 100 of them in multiples.

The stopgap multi-quarter pipeline was then run again on the earlier
set of KOIs, after removing the sections of the light curves affected
by the previously identified transits, to look for additional
candidates.  This process was iterated until no more candidates were
found.  This effort identified more than 100 new candidates in
multiple systems, in addition to the candidates that had been
identified previously using the quarter-by-quarter analysis.

\section{FALSE POSITIVES}

KOIs are reviewed from time to time by the \kep\ team, to determine
which ones should be prioritized for additional follow-up observations
of various types, and which ones are likely false positives that can
be retired to the inactive list.  In the paper summarizing the
characteristics of the planet candidates identified in the first four
months of \kep\ data, \citet{Borucki:11b} present a list in their Table
4 of 498 KOIs that had been identified as false positives and were no
longer considered to be viable planet candidates.  In Table 1 we
summarize the number of false positives compared to the number of
surviving candidates among the KOIs, with a separate accounting for
the singles and multiples.

More than half of the false positives (59\%) resulted from an ``active
pixel offset'' (APO).  This test uses a difference image analysis to
show that during transit-like events the image is significantly
displaced from the target position and is star-like, indicating
contamination by a faint background eclipsing binary or by the
wings of the PSF from a nearby bright star encroaching on the edge of
the target aperture. Most of the remaining false positives also
involved eclipsing binaries, and were identified either by features in
the \kep\ light curves, such as secondary eclipses or ellipsoidal
variations or eclipse-time variations (34\%), or by large variations
observed in the radial velocities (5\%).  Eleven of the early KOIs
were judged to be photometric false alarms, based on additional data
from subsequent quarters.

The difference in the rate of false positives for singles compared to
multiples is striking; for the singles the rate is 37 percent (486
false positives compared to 827 survivors), but for the multiples the
rate is only 3 percent (12 false positives in 6 systems, compared to
408 surviving planets in 164 systems).  This large difference is
expected, because the APOs for the singles are the result of chance
alignments of eclipsing binaries with the full target list of
nominally 150,000 stars, while the APOs for the multiples come from
chance alignments with systems that already show transit-like events.
Lumping together all the false positives among singles due to
eclipsing binaries gives a rate of (288+164+23)/150,000 = 0.0032.
Assuming that the KOIs have been drawn from the same parent population
as all 150,000 targets, the expected number of doubles involving a
planet and a false positive due to an eclipsing binary is roughly $827
\times 0.0032 = 2.6$, while the number involving two eclipsing
binaries and no planets is roughly $486 \times 0.0032 = 1.5$.  The key
assumptions here are that the probability that a target image is
contaminated by an accidental alignment with a background eclipsing
binary is the same, whether the target already shows a transit-like
event or not, and that the false-positive probability for a KOI is
independent of the depth of the transit-like event in its light curve
(multiples mostly show shallower dips).  By the same line of argument,
the number involving two planets and an eclipsing binary is only $115
\times 0.0032 = 0.4$.  Higher order coincidences are correspondingly
less likely.  The probability that a double consists of an accidental
alignment of two unrelated singles seems less likely than the two
eclipsing binary case, because eclipsing binaries are more common than
singles in the \kep\ sample by a factor of almost three
\citep{Prsa:11}.  These numbers are summarized in Table 1.

These rough estimates of the expected rates of false positives among
the multiples are preliminary, because the vetting effort is
still unfinished.  Furthermore, some types of false postives, such as
hierarchical triples, are extremely difficult to identify, especially
for shallow events.  Thus we expect that there are still false
positives lurking among both the singles and the multiples.
Nevertheless, in general it must be true that false positives due to
chance alignments must be much less common among multiples than
singles, and even for singles the rate of residual false positives
among the vetted candidates may be as low as 5 or 10 percent
\citep{Morton:11}.

\section{PLANET RADIUS VS ORBITAL PERIOD}

The process of fitting transit models to KOI light curves delivers two
primary observable characteristics of the candidate planet: the
planetary radius, \rpl\ (where we have adopted the KIC value for the
stellar radius), and the orbital period, $P$.  The plot of these two
quantities against each other is shown in Figure 1, where the active
KOIs in multiples are blue and the singles are red.  Single planets
come in all sizes, but there are relatively few giant planets in
transiting multiples.  The pile-up of giant planets near 3 days is
obvious, with no corresponding pile-up of planets, either large or
small, among the multiples.  The detection limit is especially clear
in the lower right corner of the upper panel, where period is
logarithmic. The edge of the distribution of detected candidates has a
slope of 1/3 as expected (the total number of data points during
transits varies nominally as $P^{-2/3}$).

The distributions of planet radius versus period are shown more
quantitatively by the histograms in Figure 2, where we have collapsed
Figure 1 onto the \rpl\ and $\log P$ axes in the upper and lower
panels, respectively.  The vertical scales for the singles and
multiples have been normalized so that they both have the same area
under their histograms.  Planets smaller than Neptune dominate both
samples, but more so for the multiples; $69^{+2}_{-3}$\ percent for
the singles and $86^{+2}_{-5}$ percent for the multiples.  The error
estimates for these percentages only consider Poisson noise and do not
include any contribution from uncertainties in \rpl. The difference in
the radius distributions between the singles and the multiples is
highly significant; the K-S test gives a probability of
$2\times10^{-10}$ that they are drawn from the same parent
distribution.

The period distributions for singles and multiples are quite similar.
To the eye there may appear to be a slight shift to shorter periods
for the singles, but the significance of this difference is not
supported by the K-S test, which reports a probability of 10 percent
that such differences could occur by chance.  

Figure 3 compares the number of singles versus the number of {\em
systems} that are multiples, as a function of effective temperature.
Because nearly all of the host stars are on or near the main sequence,
effective temperature is a reasonable proxy for host-star mass.  The
K-S test reports that the difference between the two distributions is
marginally significant, with a probability of 0.008 and $D$ value of
0.1.  It appears that singles may be more common than multiples around
the hotter, more massive stars, while the multiples are more common
than singles around the cooler, less massive stars.  This might be
related to the tendency for close-in giant planets to be less common
around low-mass stars \citep{Johnson:10}, and/or to the tendency for
small planets in the \rpl\ range 2 to 4 Earth radii to be much more
common around cool stars in the \teff\ range 3600 to 4100 K
\citep{Howard:11}.  Note that this is only a comparison of singles to
multiples, and no corrections have been made, either for the relative
number of targets as a function of \teff\, or for the probability of
detection.  

\section{DISCUSSION}

The fraction of single planet candidates that are smaller than Neptune
is $69^{+2}_{-3}$\ percent (569/827).  The fraction of multiple {\em
systems} with no planets larger than Neptune is $78^{+4}_{-7}$ percent
(133/170).  Thus, systems with multiple transiting planets are less
likely to include a transiting giant planet.  If the comparison is
restricted to short period planets ($P < 10$ days), the difference is
particularly striking: the fraction of short-period single candidates
that are smaller than Neptune is $69^{+3}_{-4}$ percent (279/405),
while the fraction of multiple systems that contain at least one
short-period planet (117 systems) but no short-period planets larger than
Neptune is $96^{+2}_{-9}$ percent (112/117).  One possible
interpretation is that a close-in giant planet can stir up the orbits
of other inner planets in its system, while a system of small planets
is more likely to preserve the flatness of the disk from which it
formed.  This picture is supported by determinations of the spin/orbit
alignment for transiting giant planets using the Rossiter-McLaughlin
effect. The orbits of some giants are well aligned with the rotation
of their host star, while others show significant orbital
inclinations, including even retrograde orbits
\citep{Winn:10,Triaud:10}.  Thus there is good evidence that some
systems have been disrupted from their presumably flat initial
configuration.  On the other hand, there is good evidence that
close-in giant planets do not always disrupt or prevent the formation
of systems with multiple planets.  Radial-velocity surveys show that
about 25\% of the giant planets are accompanied by companions with
smaller minimum masses \citep{Wright:09,Schneider:11}, and the actual
fraction may be much higher due to the radial-velocity detection limit
for small planets.

We observe the rate of false positives due to eclipsing binaries to be
much smaller for multiples than singles.  This is expected, because
the number of candidates that show a candidate planet or false
positive is much smaller than the full list of approximately 150,000
targets. Thus, the probability that an eclipsing binary contaminates
the light of a multiple is much smaller than for a single.
\citet{Lissauer:11b} present some independent evidence that many of
the multiples must be systems of planets, in particular the common
occurrence of periods near mean motion resonance.  This reinforces the
impression that KOIs in multiples are very likely to be planets.

Why hasn't CoRoT announced any multiples yet? {\em Kepler's} better
photometric precision and longer time series both contribute to the
detection of smaller planets with longer periods, as is needed to
discover flat systems.  Actually, CoRoT may have come very close to
detecting a multiple transiting system, namely CoRoT-7.  The
transiting planet in this system, CoRoT-7b, is the smallest discovered
so far by CoRoT (see Figure 1), but the orbit has a rather extreme
impact parameter.  Thus additional planets in the system, such as the
proposed second planet CoRoT-7c \citep{Queloz:09}, would be less
likely to transit also.

Transit time variations for planets in multiple systems promise to be
an important tool for constraining the masses of planets that are too
small to be detected with current radial-velocity techniques.  These
constraints improve with longer time series, which is a good argument
for extending the \kep\ mission and continuing to monitor the most
promising multiple systems.  This approach may be able to confirm
rocky planets in the Habitable Zones of \kep\ targets \citep{Ford:11}.

We thank the entire \kep\ team for all the hard work that has made
these results possible.  Funding for this Discovery Mission is
provided by NASA's Science Mission Directorate.  We give special
thanks to the anonymous referee for insightful and timely feedback.

{\it Facilities:} \facility{The \kep\ Mission}


\begin{deluxetable}{ccr}
\tablewidth{0pc}
\tablecaption{Census of False Positives}
\tablehead{
\colhead{Number}                        &
\colhead{Description}                   &
\colhead{Fraction}                      
}
\startdata

1733 & KOIs identified for 1489 targets     &                   \\
1235 & Survivor KOIs for 997 targets        &$ 1235/1733=0.713 $\\
827  & Survivors in singles                 &$ 827/1235=0.670  $\\
408  & Survivors in multiples               &$ 408/1235=0.330  $\\
     & 170 multiple systems                 &$ 170/997=0.171   $\\
     & 115 systems with 2 KOIs              &$ 230/408=0.563   $\\
     & 45 systems with 3 KOIs               &$ 135/408=0.330   $\\
     & 8 systems with 4 KOIs                &$  32/408=0.078   $\\
     & 1 system~ with 5 KOIs                &$   1/408=0.002   $\\
     & 1 system~ with 6 KOIs                &$   1/408=0.002   $\\
\\
498  & False positives                      &$ 498/1733=0.288  $\\
486  & False positives in singles           &$ 486/498=0.976   $\\
     & Rate per KOIs in singles             &$ 486/(486+827)=0.370  $\\
288  & Active pixel offsets (APO)           &$ 288/486=0.592   $\\
164  & Light curve evidence                 &$ 164/486=0.337   $\\
23   & Spectroscopic binaries               &$ 23/486=0.047    $\\
11   & Photometric false alarms             &$ 11/486=0.023    $\\
\\
12   & False positives in 6 systems         &$ 12/498=0.024    $\\
     & Rate per KOIs in multiples           &$ 12/420=0.029    $\\
5    & APOs in 3 systems                    &$ 3/170=0.017     $\\
5    & Light curve evidence                 &$ 5/420=0.012     $\\
2    & In one spectroscopic binary          &$ 1/170=0.006     $\\

\enddata
\end{deluxetable}

\begin{figure}
\plotone{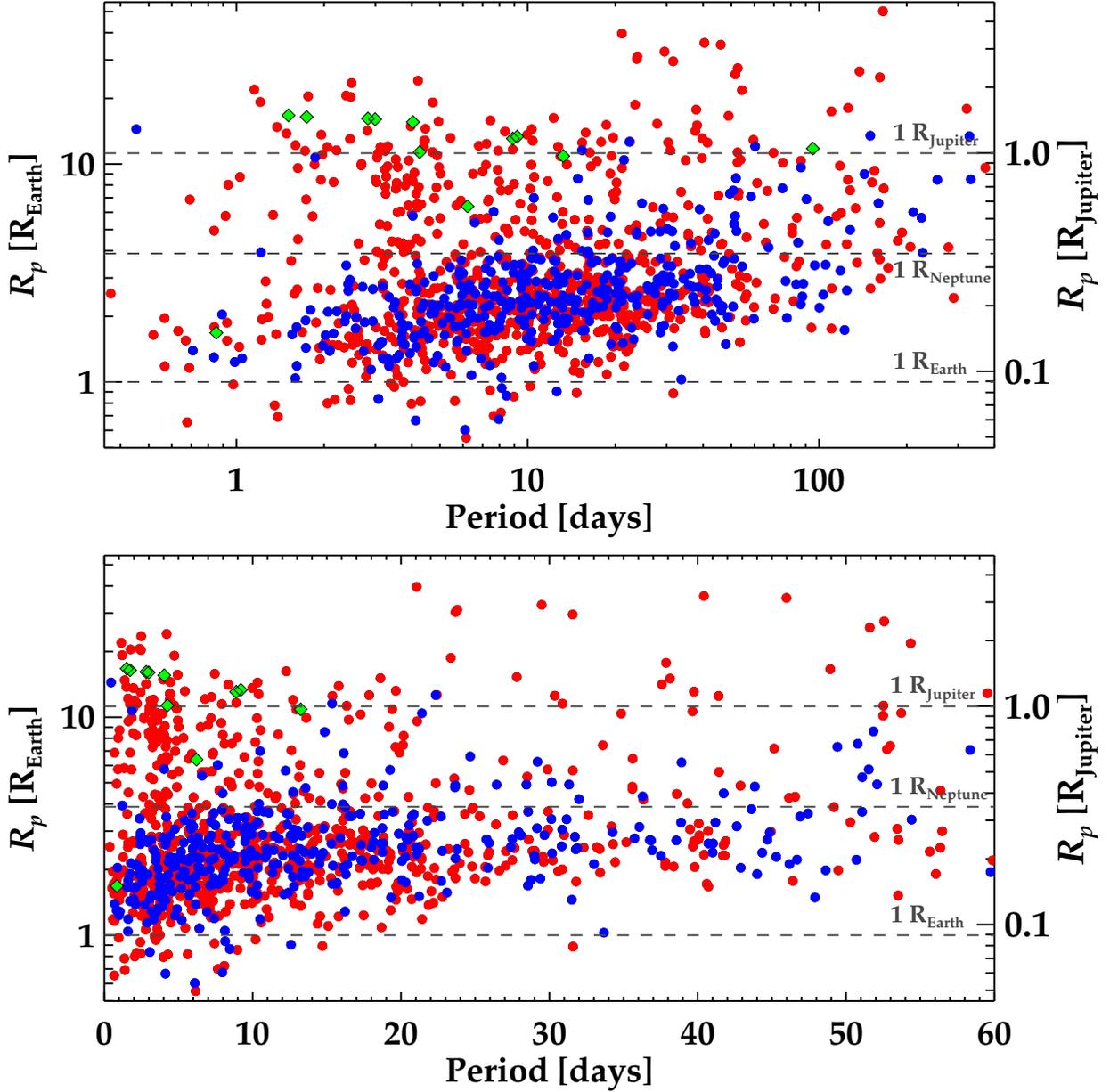}
\caption{Planet radius versus orbital period. The bottom panel shows
periods shorter than 60 days using a linear scale, the top panel
uses a log scale to show the entire range of periods.  Planet
candidates in singles are plotted in red, and in blue for those in
multiples.  \kep\ finds very few giant planets in systems of multiple
planets.  This conclusion is not affected by the rather large upper
limit that was adopted by \citet{Borucki:11b} for \rpl.  The published
CoRoT planets are plotted in green. Only one of these, CoRoT-7b, is
smaller than Neptune, and the radial-velocity observations suggest
that it may have a non-transiting companion \citep{Queloz:09}}
\end{figure}


\begin{figure}
\plotone{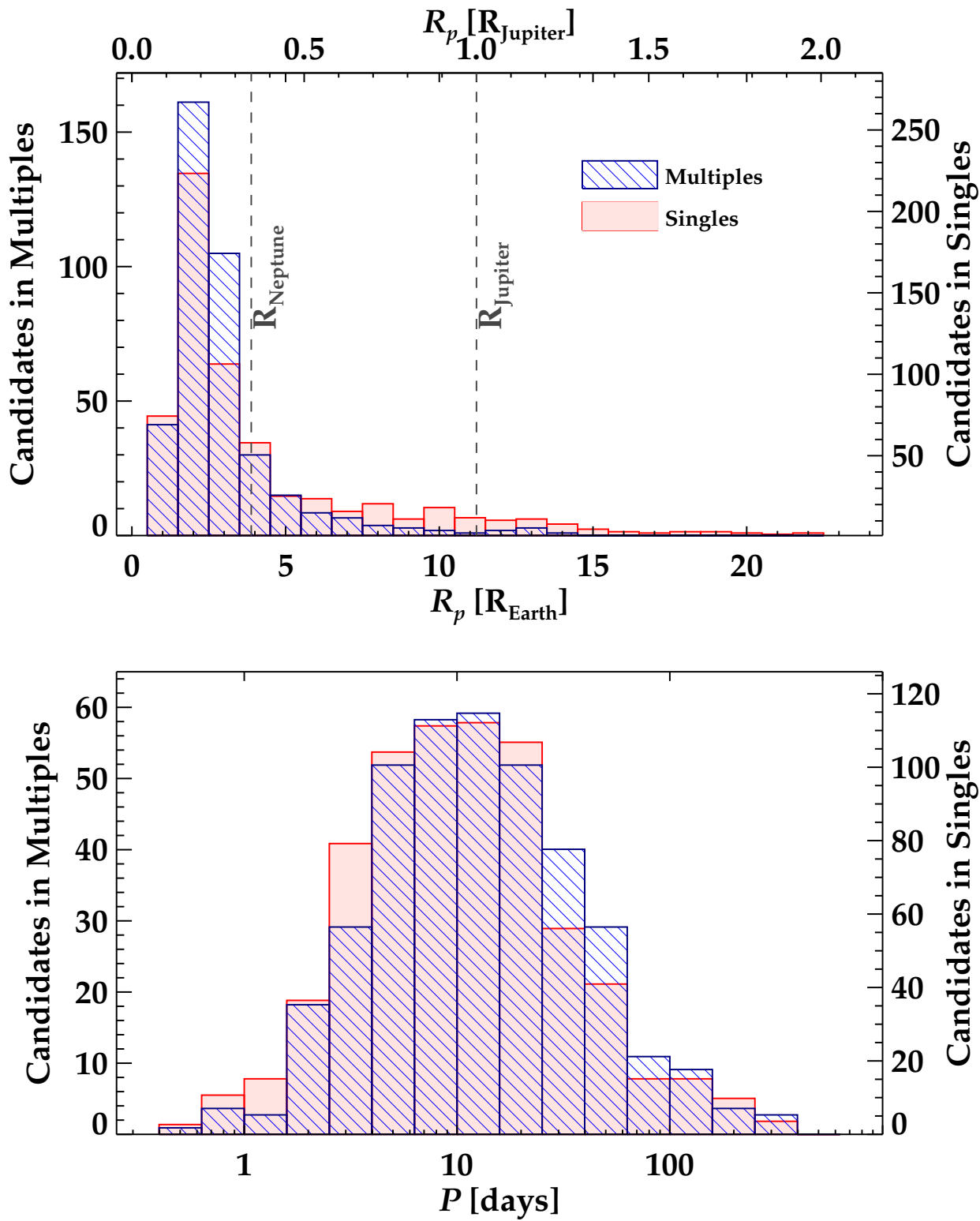}
\caption{Histograms for the number of planet candidates versus planetary radius
and period.  Singles are shown in shaded red, multiples in
cross-hatched blue.  The vertical scales for the singles and multiples have
been normalized so that they both have the same area under their
histogram.  Planets smaller than Neptune dominate both samples, but
more so for the multiples; 69\% for the singles and 86\% for the
multiples.}
\end{figure}


\begin{figure}
\plotone{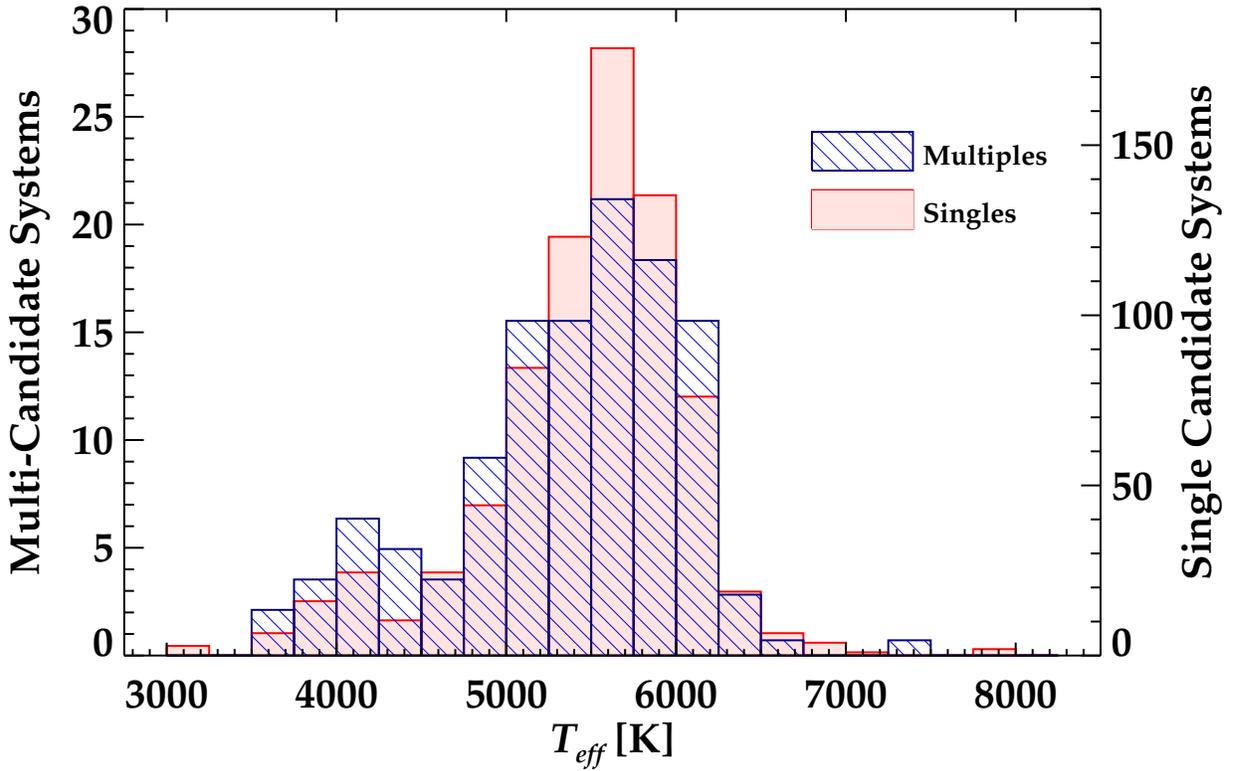}
\caption{Number of {\em systems} versus effective temperature of the
host star, which is a proxy for the stellar mass.  Single planets are
relatively more common around the hotter stars, multiple planets are
more common around the cooler stars. No corrections have been made for
the relative number of targets as a function of \teff, or for the
probability of detection.}
\end{figure}


\begin{thebibliography}{}

\bibitem[Borucki et al.(2011a)]{Borucki:11a}
Borucki, W. J., et al.~2011a.
\apj, 728, 117

\bibitem[Borucki et al.(2011b)]{Borucki:11b}
Borucki, W. J., et al.~2011b.
Submitted to \apj\ (arXiv:1102.0541)

\bibitem[Brown et al.(2011)]{Brown:11}
Brown, T. M., Latham, D. W., Everett, M. R., \& Esquerdo, G. A.~2011.
Submitted to \aj\ (arXiv1102.0342)

\bibitem[Ford et al.(2011)]{Ford:11}
Ford, E. B., et al.~2011.
Submitted to \apj\ (arXiv:1102.0544)

\bibitem[Holman \& Murray(2005)]{Holman:05}
Holman, M. J., \& Murray, N. W.~2005.
Science, 307, 1288 

\bibitem[Holman et al.(2010)]{Holman:10}
Holman, M. J., et al.~2010.
Science, 330, 51

\bibitem[Howard et al.(2011)]{Howard:11}
Howard, A. W., et al.~2011.
Submitted to \apj\ (arXiv1103.2541)

\bibitem[Jenkins et al.(2010)]{Jenkins:10}
Jenkins, J.~M., et al.~ 2010.
SPIE, 7740, 10

\bibitem[Johnson et al.(2010)]{Johnson:10}
Johnson, J. A., et al.~2010.
\pasp, 122, 905

\bibitem[Koch \& Borucki(1996)]{Koch:96} 
Koch, D., \& Borucki, W.~1996. 
In the First International Conference on Circumstellar Habitable
Zones, ed. by L.~R.~Doyle, (Travis House Pub., Menlo Park) p. 229

\bibitem[Kov\'acs, Zucker, \& Mazeh(2002)]{Kovacs:02}
Kov\'acs, G., Zucker, S., \& Mazeh, T.~2002. 
\aap, 391, 369

\bibitem[Lissauer et al.(2011a)]{Lissauer:11a}
Lissauer, J. J., et al.~2011a.
Nature, 470, 53

\bibitem[Lissauer et al.(2011b)]{Lissauer:11b}
Lissauer, J. J., et al.~2011b.
Submitted to \apj\ (arXiv:1102.0543)

\bibitem[Morton \& Johnson(2011)]{Morton:11}
Morton, T. D., \& Johnson, J. A.~2011.
Submitted to \apj\ (arXiv:1101.5630)

\bibitem[Prsa et al.(2011)]{Prsa:11}
Pr\v{s}a, A. et al.~2011
\aj, 141, 83

\bibitem[Queloz et al.(2009)]{Queloz:09}
Queloz, D., et al.~2009.
\aap, 506, 303

\bibitem[Ragozzine \& Holman.(2011)]{Ragozzine:11}
Ragozzine et al.~2011.
Submitted to \apj\ (arXiv:1006.3727)

\bibitem[Schneider(2011)]{Schneider:11}
Schneider, J.~2011.
http://exoplanet.eu/

\bibitem[Steffen et al.(2010)]{Steffen:10}
Steffen, J. H., et al.~2010. 
\apj, 725, 1226

\bibitem[Torres et al.(2011)]{Torres:11}
Torres, G., et al.~2011.
\apj, 727, 24

\bibitem[Triaud et al.(2010)]{Triaud:10}
Triaud, A.~H.~M.J., et al.~2010.
\aap, 524, 25

\bibitem[Winn et al.(2010)]{Winn:10}
Winn, J. N., et al.~2010.
\apj, 718, 145

\bibitem[Wright et al.(2009)]{Wright:09}
Wright, J. T., et al.~2009.
\apj, 693, 1084

\bibitem[Wu et al.(2010)]{Wu:10}
Wu, H.~, et al.~2010.
SPIE, 7740, 42

\end{thebibliography}
\end{document}